\documentstyle[sprocl]{article}
\newcommand{\PSbox}[3]{\mbox{\rule{0in}{#3}\includegraphics{#1}\hspace{#2}}}
\def\Journal#1#2#3#4{{#1} {\bf #2}, #3 (#4)}
\def\NPB{{\em Nucl. Phys.} B}

\begin{document}
\vspace*{-2cm}
{\hbox to\hsize{hep-ph/9612481 \hfill  MIT-CTP-2596}}
\vspace*{1cm}

\title{SIGNATURES OF PSEUDO-GOLDSTONE BOSONS IN TECHNICOLOR THEORIES
WITH THE GIM MECHANISM}

\author{ WITOLD SKIBA }
\address{Center for Theoretical Physics, Massachusetts Institute of 
Technology, \\ Cambridge, MA 02139, USA}
%%%%%%%%%%%%%%%%%%%%%%%%%%%%%%%%%%%%%%%%%%%%%%%%%%%%%%%%%%%%%%
% You may repeat \author \address as often as necessary      %
%%%%%%%%%%%%%%%%%%%%%%%%%%%%%%%%%%%%%%%%%%%%%%%%%%%%%%%%%%%%%%

\maketitle\abstracts{
Technicolor models with the GIM mechanism are successful solutions
to common problems of technicolor model building. These models have
a rich set of new particles and interactions. We present the spectrum
of the lightest exotic particles in such models and their signatures
in present and future collider experiments.}

One of the few phenomenologically viable technicolor (TC) scenarios
are models that incorporate the GIM mechanism~\cite{LR,HG} (TC-GIM).
A large value of the top quark mass requires
that the scale of extended-technicolor interactions is not much larger
than 1 TeV. In many TC scenarios, this leads to unacceptably large
flavor-changing neutral currents. Viable models should not violate
the custodial $SU(2)$ symmetry too strongly, in order to successfully
predict the value of the $\rho$ parameter.

In TC-GIM models, extended-technicolor interactions are separated into
three sectors corresponding to left-handed quarks and leptons, right-handed
up sector and right-handed down sector. Due to this construction, TC-GIM
theories possess a large GIM-like flavor symmetry, which guarantees
that the flavor-changing neutral currents are sufficiently small.
These models have only one electroweak doublet of technifermion fields,
so their contribution to the $S$ parameter is small. The custodial symmetry
is preserved to a sufficient degree.

In order to cancel anomalies, several fermion families beyond the ones
in the Standard Model are needed in TC-GIM models. These fermions are
confined by new gauge interactions. Below the confinement scale, the
observable particles are pseudo-Goldstone bosons (PGBs) associated with
spontaneous breaking of approximate chiral symmetries of these new
interactions. We will describe the spectrum and the properties of these
PGBs in a particular, so-called high-scale TC-GIM model~\cite{LR,me}.
However, most of the properties are similar for the whole class of TC-GIM
models. Note that in generic TC models PGBs are associated with chiral
symmetry breaking by the technicolor gauge group. Here, they are due
to extra confining interactions.

There are two analogous sectors of PGBs in TC-GIM theories: one associated
with the up quarks and neutral leptons, another with down quarks and charged
leptons. We will focus on the down-type sector. The PGBs arise at a
characteristic scale of confining interactions $f_{S-1}$. This scale
governs the masses of the PGBs and the strength of their couplings.
The scale $f_{S-1}$ is an undetermined parameter of the TC-GIM models,
it can be between 65 GeV and 1 TeV\@. The upper limit comes from
consistency of the TC-GIM scenario, while the lower one reflects the fact
that leptoquarks, if exist, are heavier than 130 GeV\@.
The spectrum of PGBs and their most important decay modes are presented in
Table~\ref{tab:spectrum}. Note that the heaviest particles are color-octet
and color-triplet ones. Their masses are proportional to the scale $f_{S-1}$.
The lightest, electric and color neutral, particles have masses independent
of the scale of confining interactions in the lowest-order of the
chiral perturbation theory.
\begin{table}[t]\caption{The spectrum of the pseudo-Goldstone bosons
and their decay modes.\label{tab:spectrum}}
\vspace{0.4cm}
 \begin{center}
   \begin{tabular}{|lcccc|}
     \hline 
           Particle &$SU(3)_{color}$& Charge & Mass [GeV] & Decay modes \\ 
     \hline
        $\theta^i_a$& 8             & 0      & 178 ($f/65$)& $q \bar{q}$  \\ 
        $\theta_a$  & 8             & 0      & 178 ($f/65$)& $gg, \gamma g, 
                                                             q\bar{q}$   \\ 
        $T^i_c, T_c$& 3             & 2/3    & 134 ($f/65$)& $q \bar{l}$  \\ 
        $\Pi^i, P^i$& 1             & 0      & 0.1--5     & $q \bar{q},
                                                            l \bar{l}$   \\ 
        $P^0$       & 1             & 0      & 1          & $gg, q\bar{q},
                                                             l\bar{l}$ \\
     \hline
   \end{tabular}
   \vspace{-5pt}
 \end{center}
\end{table}

We now turn to description of the PGBs couplings. The PGBs couple to
fermion pairs via derivative coupling. The effective coupling can be
described by
\begin{equation}
  \label{eq:coupling}
  \left( \frac{f_{S-1}}{f_{ETC}} \right)^2
  \bar{\psi_L} \gamma^\mu (\partial_\mu \Sigma) \Sigma^\dagger \psi_L
  \; {\rm and} \;
  \left( \frac{f_{S-1}}{f_{ETC}} \right)^2
  \bar{d_R} \gamma^\mu (\partial_\mu \Sigma) \Sigma^\dagger d_R,
\end{equation}
where $\Sigma = \exp{(2 i T_a \pi^a / f_{S-1})}$ is the non-linear
representation of the PGBs, while $f_{ETC}$ is the characteristic scale of
the extended-technicolor interactions. These couplings are quite different
from generic TC models, where the fermion-PGB couplings are non-derivative
and proportional to the mass of a PGB. The coupling of PGBs to gauge bosons
is described by the minimal coupling imposed by the gauge invariance. It is
important to note that the constituents of the PGBs do not transform under the
electroweak $SU(2)$, so they do not couple to $W^\pm$. However, the charged
PGBs couple to the $Z^0$ boson due to the hypercharge interactions.
Also, the $P^0$ and the $\theta^a$ bosons have anomalous couplings to
gauge boson pairs: either gluons or photons ($Z^0$). The amplitude
of the anomalous coupling is inversely proportional to the scale $f_{S-1}$.

\begin{figure} 
 \PSbox{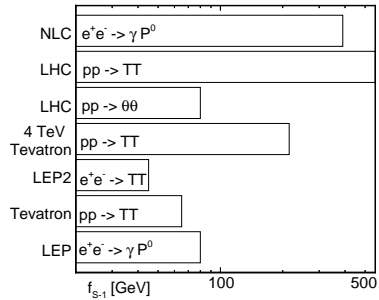 hscale=30 vscale=30 hoffset=60 
                             voffset=-100}{11.cm}{3.4cm} 
\caption{The potential of probing the scale $f_{S-1}$ by collider experiments.
 \label{fig:colliders}}
 \vspace{-5pt}
\end{figure}
A very light $P^0$ boson is a new feature of the TC-GIM scenario. This particle
can be produced in $e^+e^-$ collisions via its anomalous couplings. The
process to search for is $e^+e^- \rightarrow P^0 \gamma$, with subsequent
decays of $P^0$. Decays of $P^0$ into a small number of hadrons dominate
if $f_{S_-1}$ is smaller than 200 GeV, decays into a $\mu^+ \mu^-$ pair
dominate otherwise. At present, one can gain most information about
TC-GIM by analyzing all existing data from LEP1, the signatures are very
similar to the ones of $Z^0 \rightarrow \eta' \gamma$. LEP2 will not
provide any new information about $P^0$ due to small production cross sections.
A 500 GeV $e^+e^-$ collider collecting $50 {\rm pb}^{-1}$ of luminosity
can probe $f_{S-1}$ up to 390 GeV\@. The leptoquarks $T$ can be produced
either at $e ^+e^-$ colliders or via gluon-gluon fusion in hadron colliders.
The $ep$ colliders are not helpful in studying TC-GIM particles due to
small production cross sections. The $e ^+e^-$ experiments can discover
leptoquarks almost as heavy as the kinematic limit for their production,
since their signatures are very distinct: two isolated leptons and two
jets. New generations of hadron colliders will have a huge potential
for discovering leptoquarks. A 4 TeV machine can find leptoquarks as heavy
as 440 GeV, while a 16 TeV collider as heavy as about 1160 GeV.
Hadron colliders produce the octet bosons even more copiously than they
produce leptoquarks. However, pair-produced octet particles have four-jet
signatures that are very difficult to disentangle from the backgrounds present
in $pp$ collisions. The LHC experiments can detect octet PGBs up to 375 GeV.
%Discovery of only one type of PGBs
%would be enough to estimate the scale $f_{S-1}$, which in turn would
%enable an estimate of the masses and couplings of the remaining PGBs.
In Figure~\ref{fig:colliders} we present the potential of various colliders
to probe the scale $f_{S-1}$.

\section*{Acknowledgments} 
I would like to thank Lisa Randall for many helpful discussions.
This work is supported in part by funds provided by the U.S. Department
of Energy under cooperative research agreement DE-FC02-94ER40818.


\section*{References}
\begin{thebibliography}{9} 
\bibitem{LR} L. Randall, \Journal{\NPB}{403}{122}{1993}.
\bibitem{HG} H. Georgi, \Journal{\NPB}{416}{699}{1994}.
\bibitem{me} W. Skiba, \Journal{\NPB}{470}{84}{1996}.
%\bibitem{leptoquarks} S.~Abachi, {\it et al.\ } (D0 Collaboration), 
%\Journal{\PRL}{72}{965}{1994}; \\
%F. Abe, {\it et al.\ } (CDF Collaboration), \Journal{\PRL}{75}{1012}{1995}.

\end{thebibliography}
\end{document}